\documentclass[article]{aa}

\usepackage{graphics}
\usepackage{epsfig}
\usepackage{latexsym}
\begin{document}
\hyphenation{brems-strah-lung}
\title{The 2009 outburst of H~1743-322 as observed by RXTE}

\author{Yupeng Chen\inst{1}, Shu Zhang\inst{1}, Diego F. Torres \inst{2}, Jianmin Wang\inst{1,3},    Jian Li\inst{1},
Tipei Li\inst{1,4}, Jinlu Qu\inst{1}}

\institute{Laboratory for Particle Astrophysics, Institute of High
Energy Physics, Beijing 100049, China
\and
ICREA \& Institut de Ci\`encies de l'Espai (IEEC-CSIC), Campus UAB,
Facultat de Ci\`encies, Torre C5-parell, 2a planta, 08193 Barcelona,
Spain
\and
Theoretical Physics Center for Science Facilities (TPCSF), CAS
\and
Center for Astrophysics,Tsinghua University, Beijing 100084, China
          }

\offprints{Yupeng Chen}
\mail{chenyp@mail.ihep.ac.cn}

\date{ }


\authorrunning{Yupeng Chen et al.}

  \abstract
  {}
{Six outbursts were observed by RXTE from the black hole candidate H~1743-322 since 2003. Following ``a failed outburst" with  missing high/soft state in 2008, the most recent one occurred in 2009, {\bf showing a complete trip through all spectral transitions}. We investigate this outburst in the context of the  source outburst  history. }
{We analyze the RXTE observations of the 2009 outburst of H~1743-322, as well as the observations of the   previous five outbursts for comparison.  All available RXTE observations taken since 2003  sum up to a total exposure of  $\sim$ 1200 ks.}
{The hardness-intensity diagram (HID) shows a complete counter-clockwise q-track for the 2009 outburst and, interestingly, the track \textbf{falls in} between a huge one in 2003, with a complete  transition to high/soft state, and \textbf{that of} the failed outburst in 2008. It leaves the low/hard state but does  not reach the leftmost edge of the \textbf{overall HID. Near maximum, the X-ray spectra are described by a power law} with photon index $<$ 2.4;  the peak luminosity \textbf{during the whole outburst reached  only $\sim$ 20\% of the Eddington luminosity}. The largest disk fraction \textbf{with respect to the total luminosity is} about 0.75 for the 2009 outburst, which is close to the lower limit observed from the black hole XRB sample in their  state transitions from the intermediate  to the high/soft state.   While the lowest hardness \textbf{(6--19 keV/3--6 keV) values} in the HID  is about 0.3--0.4 for the 2009 outburst, similar to the ``failed state transition" seen in the persistent black hole XRB Cyg X-1, the timing analysis shows that a transition to the high soft state occurred. During the low/hard state of the 2009 outburst, the  inner radius of the accretion disk   is found to be closer to the central black hole and have an anti-correlation with the disk temperature. These results may be understood as the \textbf{reprocessing} of the hot corona on the \textbf{ disk's} soft X-rays, which can lead to  an underestimation of the inner radius of the accretion disk.  In the luminosity diagram of the corona \textbf{versus the disk, the tracks of the outbursts} in 2003 and 2009 cross the line which represents a roughly equal contribution to the entire emission from the thermal and the non-thermal components\textbf{;}   the track of the 2008 outburst has the turn-over \textbf{falling} on this line. This may be indicative of an  emission balance between the corona and the disk, which prevents the state transition from going further than the low/hard state.}
{} 


   \titlerunning{The State Transition of H~1743-322 in the 2009 Outburst}
   \maketitle

\section{Introduction}
From an observational point of view, the low-mass X-ray binaries (LMXB), being  the compact star  a neutron or a black hole candidate (BHC), have at least two states during their outbursts: the low/hard (LHS) and the high/soft state (HSS), a distinction based on their spectral behaviour.
%
%
Many models have been developed to understand the nature of black hole candidates (BHCs), but basically, they all assume that their spectra are generated by
two (thermal and non-thermal) components.
The differences among these models arise from  the  locations in which these components are generated
 (for reviews, see, e.g., Homan \& Belloni 2005,
Remillard \& McClintock 2006,
and Belloni 2009).
When the accretion rate  increases from a quiescent state, the inner accretion disk region evaporates to form hot plasma or an advection-dominated accretion flow  region (ADAF, Narayan et al. 1996), and the seed photons are comptonized to X-ray photons.
The LHS is more luminous than the quiescent state, and most binaries were discovered when this transition occurred (Nowak 1995; Fender et al. 2004; Belloni 2009).
It is thought that for BHCs after the LHS,  the soft X-rays (e.g., in 2--10 keV) luminosity of most outbursts still increase to more than 10\% of Eddington's (\textbf{McClintock \& Remillard  2006)}; the inner disk being hotter and closer to the BHCs  than in the LHS.
 However,  circular orbits around BHs with   radii below a certain value are unstable, and the disk  radius lower limit extends only to the innermost stable circular orbit (ISCO). For a Schwarzschild BH, the ISCO is at $3GM/c^{2}\approx3.0~M_{BH}/M_{\odot}$ km.
For the BHCs beyond  the HSS, some outbursts' luminosity can increase up to  the  Eddington luminosity, placing  it into a very high state (VHS, Miyamoto et al. 1991), where the  contribution from the soft thermal component with typical temperature $\sim$ 1 keV) and the power law component (with a photon  index $\sim$ 2.5) are comparable.

There are two main intermediate states between the LHS and HSS: the  so-called hard intermediate state (HIMS) and the soft intermediate state (SIMS), where the X-rays spectra are described by a power law with photon index 1.6--2.5. They can be  distinguished in the hardness-intensity diagrams (HIDs).
In the latter,
the evolution of BHCs forms  a ``q-track" pattern,  e.g., see the outbursts evolution of GX~339-4
(Fender et al. 2004; Belloni et al. 2005;
Homan \& Belloni 2005; Del Santo et al. 2008; Belloni 2009, for a review, see Homan \& Belloni 2005; Dunn et
al. 2010).
Both the LHS and HSS are rather stable states, which can be observed  on time scales of months or years, while the remaining states show strong and fast variations.

 Besides the spectral characteristics, another feature of BHC states is fast aperiodic variability (e.g., van der Klis 1994; Vaughan et al. 2003; McHardy et al. 2004; Uttley et al. 2005). The root mean square  (RMS), i.e.,
integrated power spectral densities (PSDs) over a broad range of frequencies, is usually treated as an indicator of  the variability amplitude. Analogous to HIDs, the hardness-rms diagrams (HRDs), where the  fractional  rms is plotted versus hardness, are useful for characterizing the aperiodic variability behavior in different states ( e.g., Gleissner et al. 2004; Belloni 2009).  The fractional rms in the LHS,   typically  $\sim$30\%, increases slowly with flux.  For the HSS, the fractionsl rms is typically $<$10\%, weaker and located at the lower right corner of the HRDs. In the latter, the intermediate states have intermediate RMS value, of course.  The evolution of RMS  values is smooth in the HRDs except for the SIMS, which is marked  by a sharp drop (Belloni et al. 2005; Fender et al. 2006).

After the initial LHS, most sources have tracks extending to the left branch in the HID, and step into  the HSS.
However, a few sources, such as  XTE J1118+480  (Remillard et al. 2000; Wilson \& McCollough 2000) and H~1743-322 (Capitanio et al. 2009), sometimes have tracks that do not reach the HSS
  with peak luminosities around $\sim$ 0.1 $L_{Edd}$: they leave the LHS and go back to quiescence  without  canonical ``q" patterns in HIDs, forming the so-called ``failed outbursts".  The outburst can move further from   the LHS to form a SIMS. The case in which the SIMS is not followed by a HSS is  referred to as a  ``failed state transition".
 Although rather rare, failed state transitions have been occasionally observed  from the  persistent bright black hole binary Cyg X-1 (Belloni et al. 1996). Cyg X-1 spends most of the time in the LHS, with occasional transients to the HSS and failed state transitions. In the latter, the hard component does not become so soft as a full-fledged HSS but only reaches the SIMS  with the X-ray spectra being described by a power law with a photon index $\sim$ 2.15 and inner disk temperature of  about 0.36 keV. Spectrally, the hard component does not become too soft and does not reach the leftmost part in the HIDs as for other transient BHC. The RMS does not go down to 10\% and has no sharp QPO (for a review, see Belloni 2009).

The black hole candidate and X-ray binary H~1743-322 (also named 1H~1741-322, XTE~1746-322, and IGR~J17464-3213)
was proposed to be a BHC based on the ``ultrasoft X-ray spectra" during outburst
(White \& Marshall 1983).
In the RXTE era before 2009, five outbursts (referred to in what follows as \#1--5, in time sequence) were detected from H~1743-322: on March 2003 (MJD 52550), September 2004 (MJD 53250), September 2005 (MJD 53600), January 2008 (54450) and October 2008 (MJD 54750).
In all transient sources for which the initial rise is observed, the start of the outburst is in the LHS.
 For the first and brightest outburst in 2003, in which the source moved to  the VHS, McClintock et al. (2009) found that the source presented many similarities with the BH binary XTE~J1550-564 at the radio and X-ray bands. In this outburst, Homan et al. (2005) detected high-frequency QPOs at 240 and 160 Hz, making this source the fourth BHC  having high-frequency QPOs (the others being GRO J1655-40, GRS 1915+105, XTE J1550-564). The outbursts \#2, 3, and 4 (see Swank 2004; Rupen et al. 2005; Capitanio et al. 2006; Kalemci et al. 2008)  are less investigated because of limited observations. The fifth outburst  was a failed outburst,   with the X-ray spectra described by a power law  with the softest photon index ($\sim$  $2.1_{-0.03}^{+0.03}$) and the lowest luminosity ($\sim$ 0.1  $L_{Edd}$). This one is the faintest outburst in the set;  it is thought to lack the high/soft state because of its low  (peak) accretion rate at that time
(Capitanio et al. 2009). Lately, H~1743-322 showed a new outburst initially detected by Swift/BAT on 2009 May 26 (Krimm et al. 2009). Radio activity was also detected by VLA in H~1743-322 on 2009 May 27 (James et al. 2009). 
H~1743-322 went back to the low/hard state on 2009 July 7, with  power law photon index dropping to 1.84 (Kalemci et al. 2009).
Renewed activity from H~1743-322 was detected by MAXI/GSC on 2009 December 27  with a  power-law photon index of 1.7, and the source was suggested to be in the LHS (Yamaoka et al. 2009).

In this paper, we carry out a systematical  analysis on the most recent outburst (\#6), for which only  some preliminary  results  were reported (see e.g. Krimm et al. 2009, and references therein), based on the most updated  RXTE  observations. We put our results in context of all the other outbursts detected from H~1743-322,  e.g.,  Parmar et al. 2003; McClintock et al. 2009; Capitanio et al. 2009; and Dunn et al. 2010).

\section{Observations and data analysis}

Public data from {\it RXTE~} (Gruber et al. 1996) on H~1743-322 covers the period between March 2003 and July 2009, and include 410 {\it RXTE}/PCA pointed observations, with the identifier (OBSID) of proposal number (PN) 80137, 80138, 80144, 90058, 90115, 90421, 91050, 91428, 93427 and 94413 in the data  archive of the  High Energy Astrophysics Science Archive Research Center (HEASARC).
These observations sum up  to $\sim$ 1200 ks of exposure time on the source, and scatter over the six outbursts.

The analysis of PCA data is here performed by using
 HEAsoft v. 6.6.  We filter the data using the standard \emph{RXTE}/PCA criteria.
 Only the PCU2 (in the 0-4 numbering scheme) has been used for the analysis, because  this PCU was 100\% on during the  observations,  we  selected the time intervals with the following constraints:  elevation angle $>10^{\circ}$, and pointing offset $<0.02^{\circ}$, where elevation is the angle above the limb of the Earth, and the pointing offset is the angular distance between the pointing and the source. The background file used in analysis of PCA data is the most recent bright-sources one found at the HEASARC website\footnote{
pca$\_$bkgd$\_$cmbrightvle$\_$eMv20051128.mdl}; the detector breakdowns have been removed.\footnote {For more information see:\\
http:$//$heasarc.gsfc.nasa.gov$/$docs$/$xte$/$recipes$/$pca$\_$breakdown.html.}
Data from cluster 1 of the HEXTE system are used to produce the  higher energy spectra.
An additional  1\% systematic error is added to the
spectra because of calibration uncertainties, if not otherwise specified.  The  X-ray
spectra are fitted with XSPEC v12.5.0 and the model parameters are
estimated  with 90$\%$ confidence level.

\section{Results}

\subsection{Lightcurves}

\begin{figure}[ptbptbptb]
\centering
 \includegraphics[angle=0, scale=0.44]{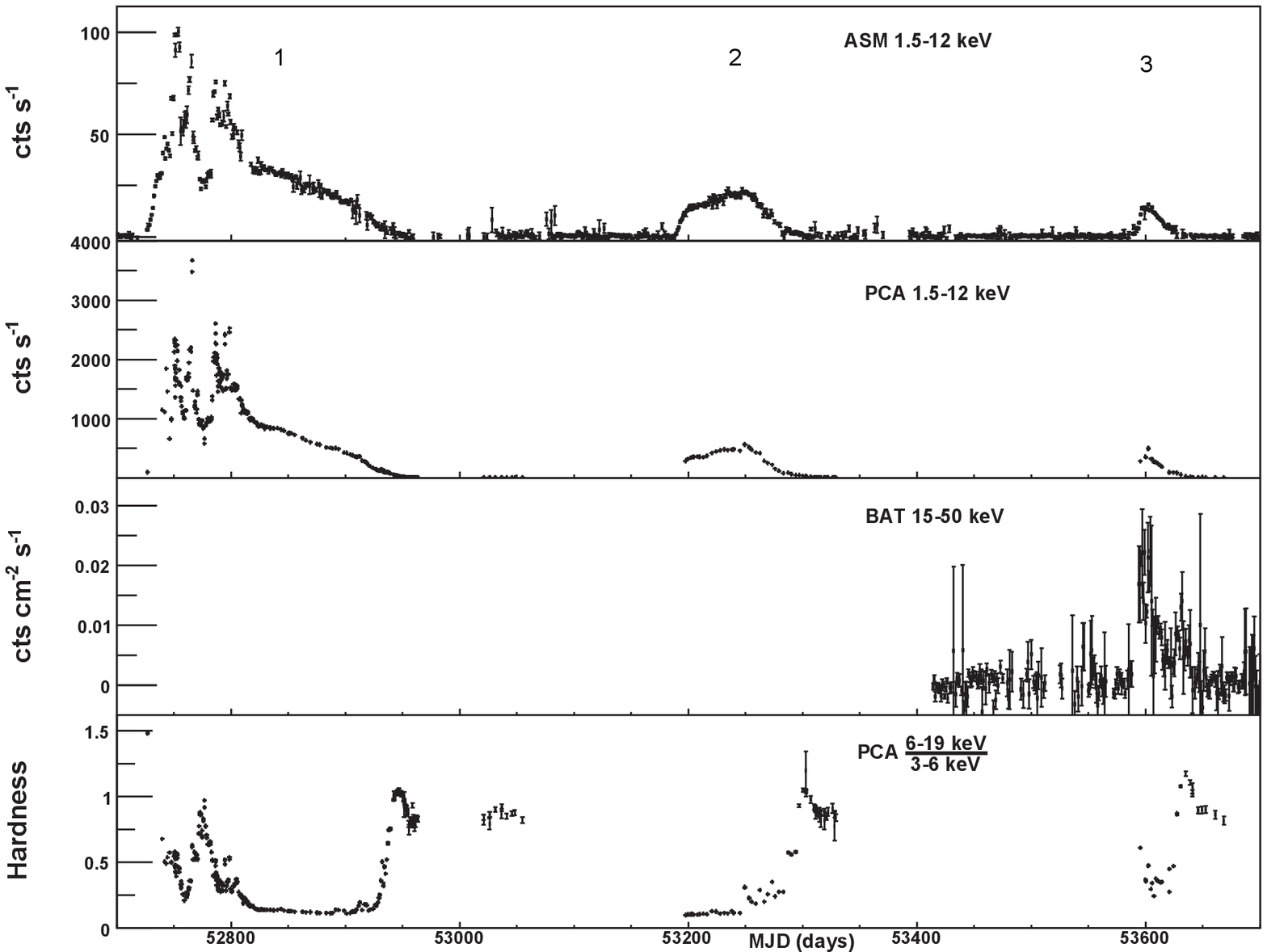}
  \includegraphics[angle=0, scale=0.44]{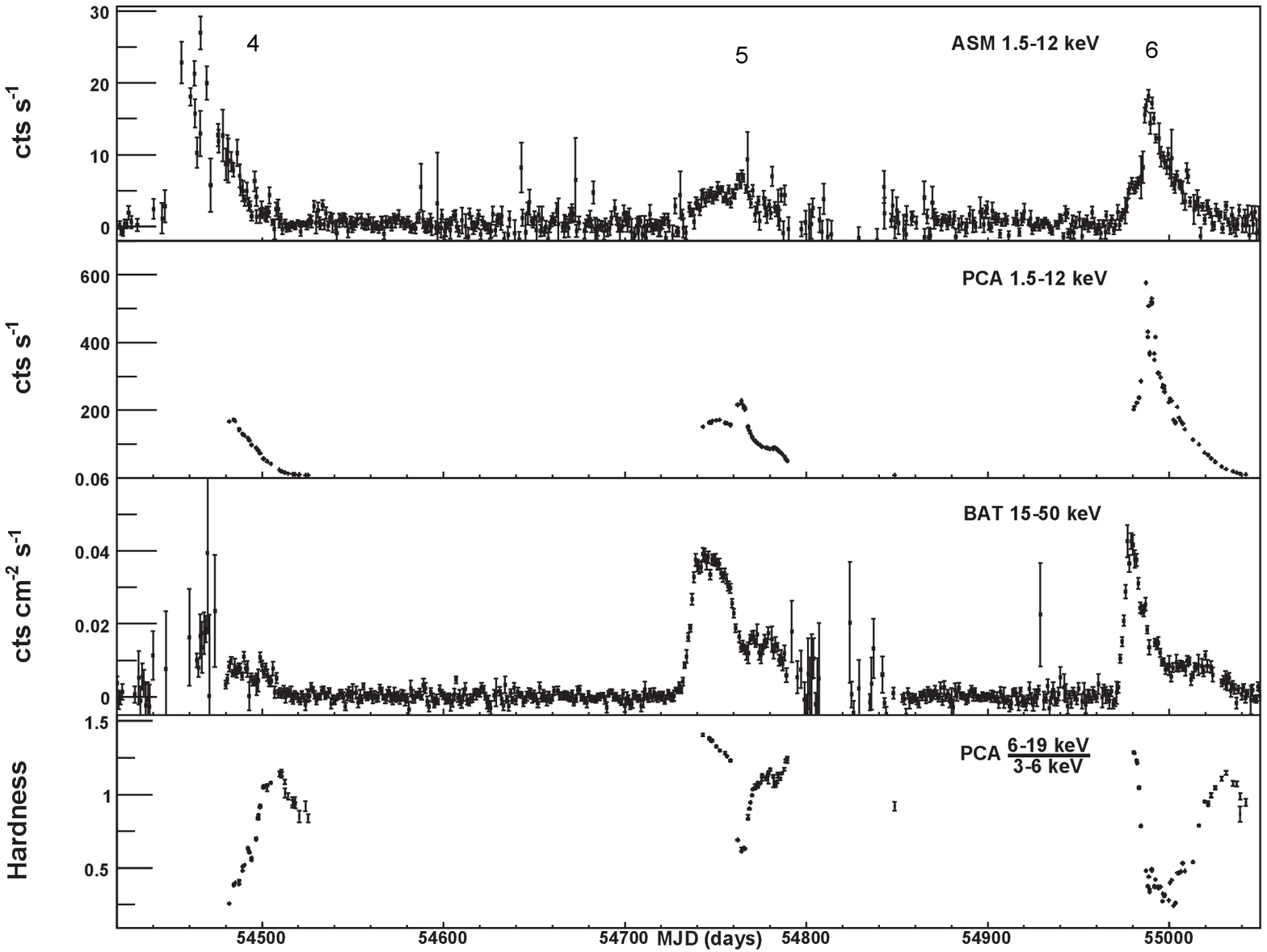}
      \caption{The ASM(1.5--12 keV), PCA(1.5--12 keV), BAT(15--50 keV) lightcurves and PCA hardness ratio (6--19 keV/3--6 keV)
      of H~1743-322 from March 2003 to July 2009. The bin size for the lightcurves  is one day, and for the hardness  is 6400 seconds.
The PCA lightcurves prior to outbursts \#6 are also available from  McClintock et al. (2009) and Capitanio et al (2009).
}
         \label{lc_flux}
\end{figure}

The lightcurves of the six outbursts (\#1--6, in time sequence) in the energy band 1.5--12 keV are shown in Fig. \ref{lc_flux}, using 6400 seconds time bins for every observation identifier (OBSID). These lightcurves  use PCA standard 2 mode data for each energy channel; the background has been removed. For comparison, we put together as well the RXTE/ASM (1.5--12 keV)\footnote{
See the lightcurves provided by
the All-Sky Monitor (ASM) team at the Kavli Institute for Astrophysics and Space Research at the Massachusetts Institute of Technology,
 http://xte.mit.edu/
}
 and Swift/BAT (15--50 keV)\footnote{See the Swift/BAT transient monitor
results provided by the Swift Team at
http://swift.gsfc.nasa.gov/docs/swift/results/transients}
daily-averaged lightcurves.
The ASM lightcurves  (Fig. \ref{lc_flux}) enclose
all the six outbursts from H~1743-322,
including the  brightest one in 2003 and  the subsequent five, fainter ones, of which the last four were caught by Swift/BAT at 15--50 keV.
The lightcurves for the outbursts \#1--5  were also analyzed by  Parmar et al. (2003), McClintock et al. (2009), Capitanio et al. (2009), and references therein. We  thus make a comparison of these with  the new outburst occurred in 2009.
The latter was well covered by RXTE/PCA for  almost the whole evolution from the LHS to the quiescent  state.
The ASM lightcurve has an outburst  at soft X-rays with a profile similar to that of the BAT lightcurve at hard X-rays, with the flux peak at soft X-rays lagging that at hard X-rays by roughly 10 days.  For the 2009 outburst, the Swift/BAT flux  increased since MJD 54977 and reached a peak on MJD 54987, followed by a decay of more than 60 days.
The hardness ratio, defined as the PCU2 flux ratio (6-19 keV)/(3-6 keV), are computed to  demonstrate at first glance   (see also Fig. \ref{lc_flux})   the spectral changes along the outburst evolution.

\subsection{Hardness-intensity diagrams}

\begin{figure}[ptbptbptb]
\centering
 \includegraphics[angle=0, scale=0.7]{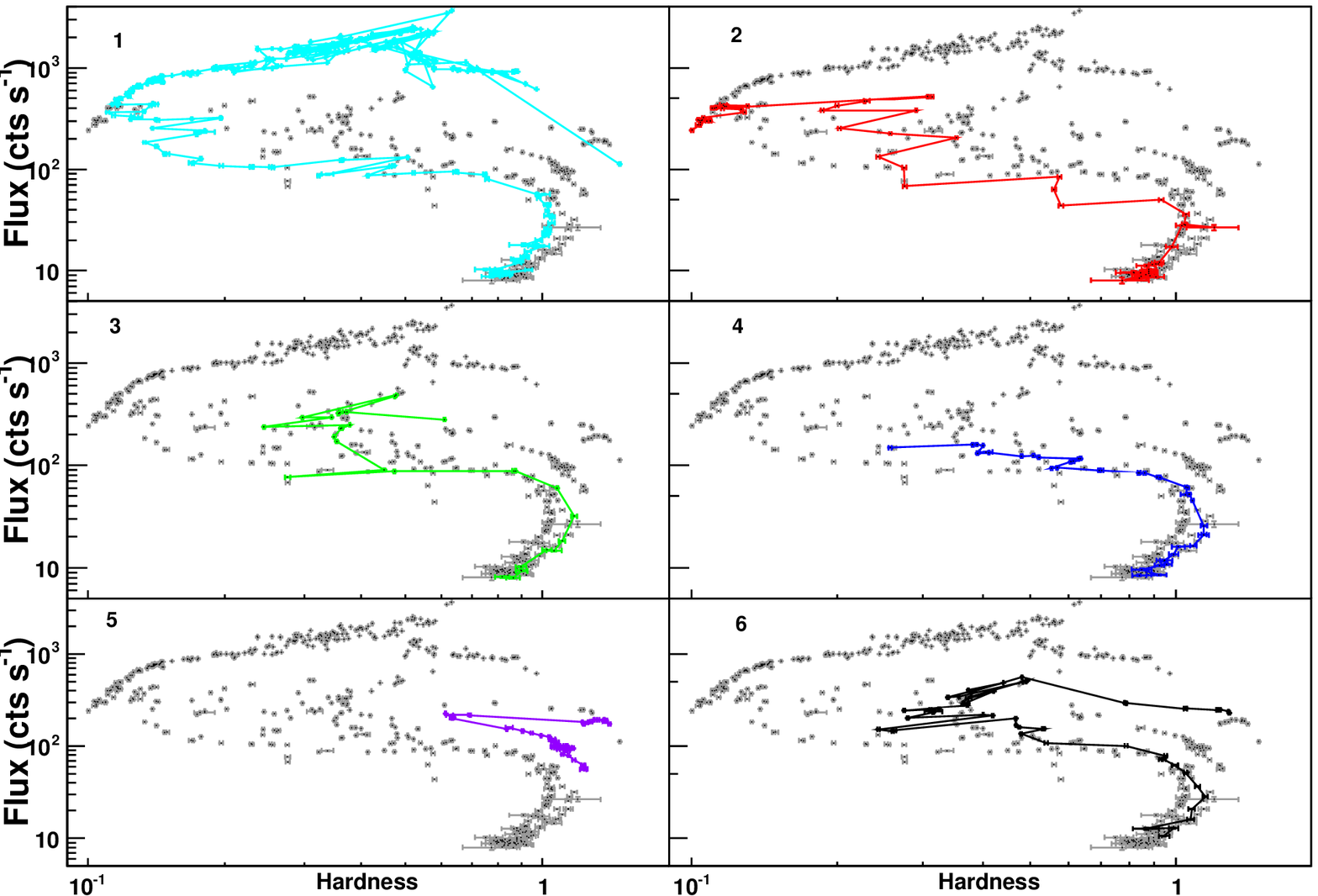}
  \includegraphics[angle=0, scale=0.3]{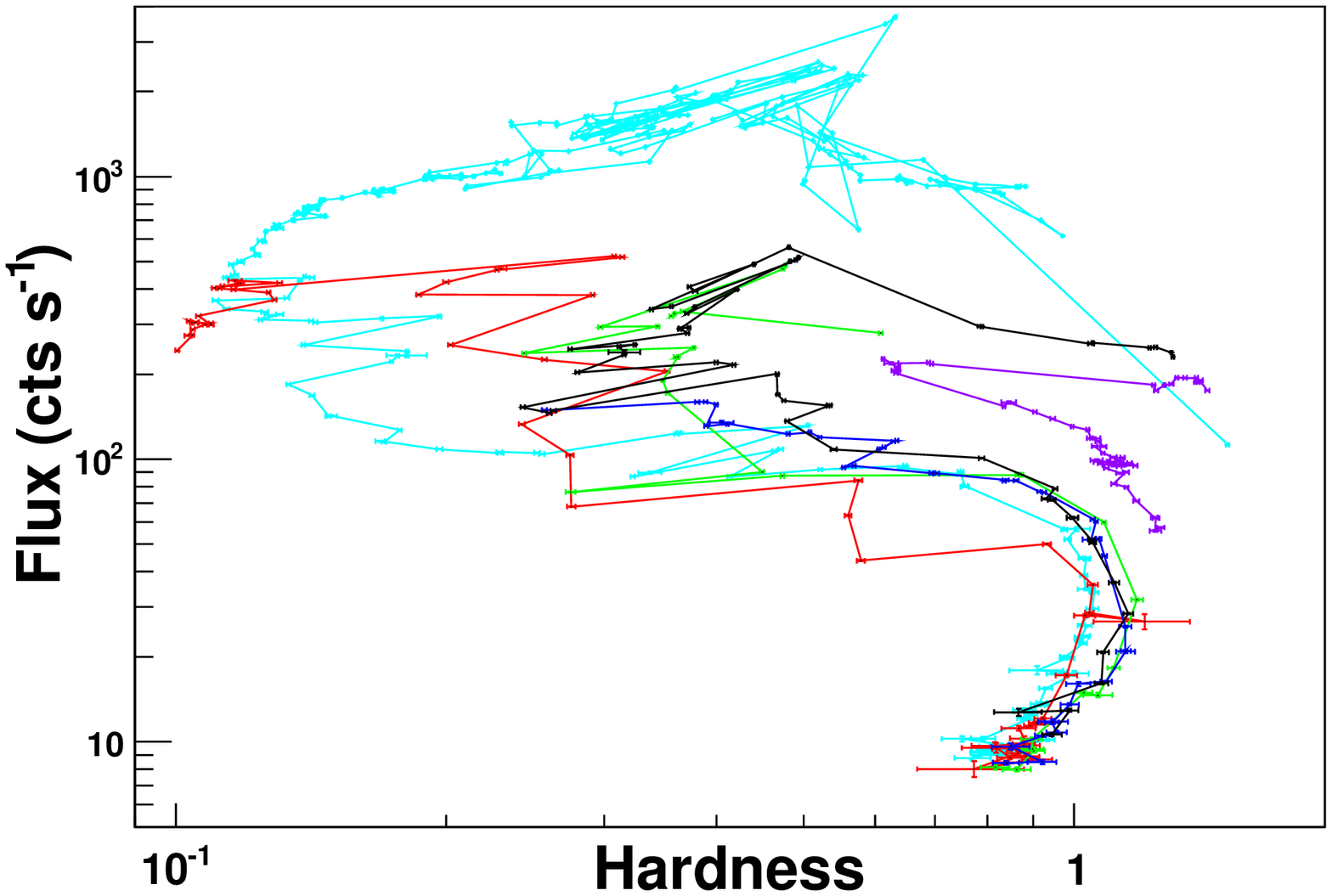}
      \caption{The hardness-intensity diagrams (HIDs) of the six outbursts observed by RXTE/PCA. The hardness is defined as the ratio of count rates between the 6--19 keV band and 3--6 keV band, as extracted from the lightcurves. The intensity is in units of  count rates in the 3--21 keV band. The gray points represent all of the six outbursts together, each outburst is plotted in color against this gray background.
For comparison, we mosaic as well the six HIDs in the bottom panel.
The time bin size for each data point  is 6400 seconds.
 Note that the HIDs were reported in Dunn et al. (2010) for the outbursts \#1--5  and in Capitanio et al. (2009) for outbursts \#4--5.
}
         \label{lc_HID}
\end{figure}

We investigate   the six outbursts  by deriving the  evolution of the  hardness  (6-19 keV)/(3-6 keV) against intensity in the 3--21 keV band (Fig. \ref{lc_HID}). The brightest  outburst, \#1, which occurred in April 2003, has a rather complete evolution track in the HID:  it shows the LHS state, the intermediate states, the HSS, and the quiescent state. For outburst \#2 (occurred in July 2004, around MJD 53200),  a partial HID track including the  HSS and the transition to quiescence were recorded by  the PCA.  For  outburst \#3 (occurred in August 2005, around MJD 53600), the initial LHS state is missed, and it has an intermediate hardness at the flux peak.  Only the decay part was observed by the PCA for outburst \#4, whereas the
whole track is present for outburst \#5.  The latter is the faintest outburst and was reported to have failed the transition beyond the LHS. The  most recent  outburst, \#6,  is the 3rd one with an almost full HID track to be recorded by the PCA. In June 2009, the outburst \#6  reached the flux peak but with a hardness located in between the outbursts \#1 and \#5.

\subsection{Spectra}

\begin{figure}[ptbptbptb]
\centering
 \includegraphics[angle=0, scale=0.9]{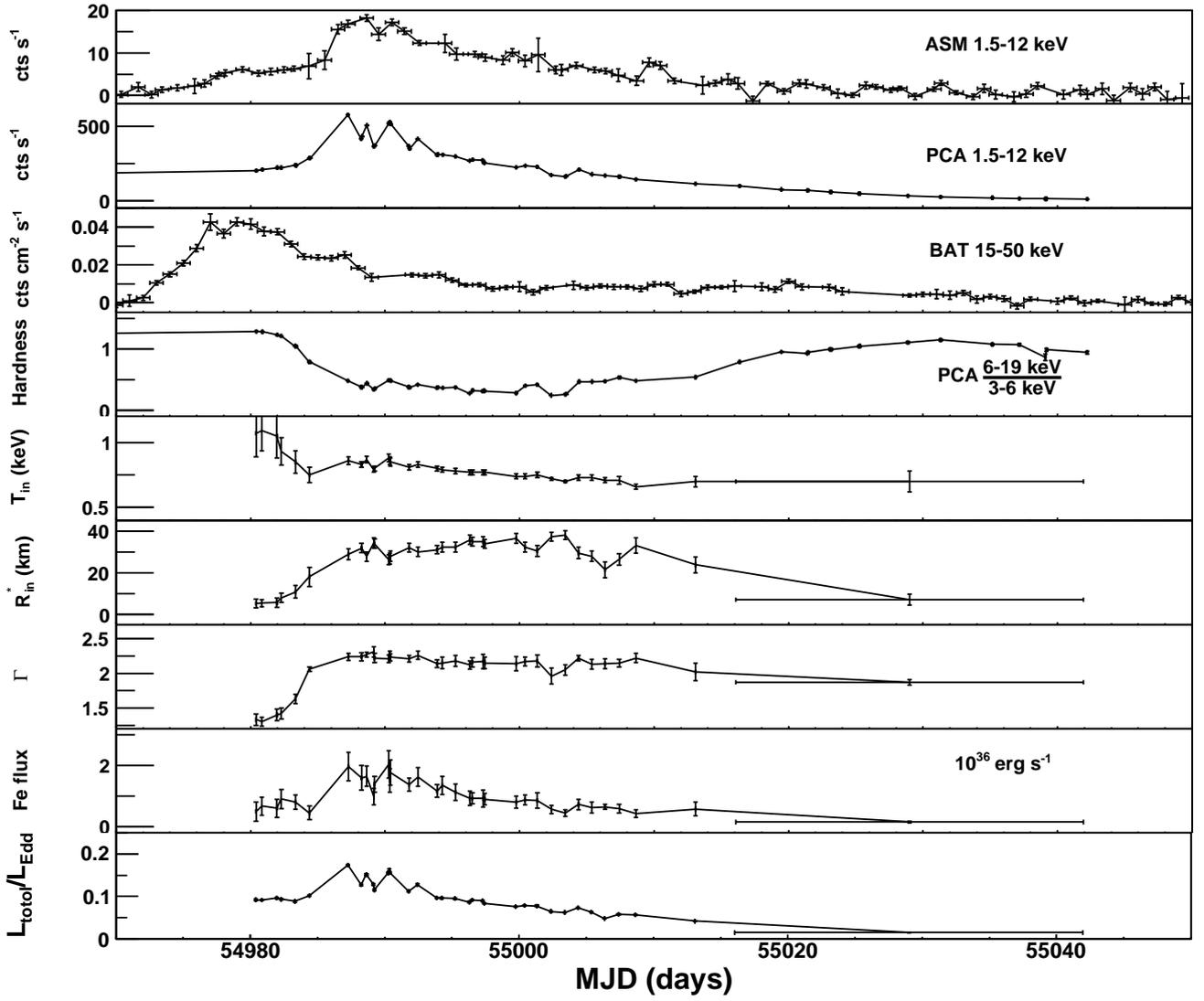}
      \caption{The time evolution of the ASM (1.5--12 keV), PCA (1.5--12 keV), and BAT (15--50 keV)  lightcurves, the PCA hardness ratio (defined using count rates corresponding to 6--19 keV and 3--6 keV),
      the inner temperature of disk $T_{in}$,
      the observed inner radius of the disk $R^{*}_{in}$,
      the power-law index $\Gamma$,
      the Fe luminosities $L_{Fe}$, and
      the total luminosities ($L_{disk}$+$L_{pow}$)
       for  the recent outburst \#6  of H~1743-322.
        The   time bin size for ASM and BAT is one day, and for the hardness curve is
        6400 seconds. The spectral bin size is the same as in Table 1.
 }
         \label{lc_spec}
\end{figure}

The spectra of the outburst \#6  are extracted for each  OBSID of the combined  PCU 2 data in the 3--20 keV band and the HEXTE cluster 1 data in the 20--100 keV band.   We take the DISKBB model plus a power-law/cutoff power-law to fit the data,
common to modeling BHC sources, under a fixed column density of 1.6$\times$10$^{22}$ atoms/cm$^2$
(given by Capitanio et al. 2009) and an iron line fixed at 6.4 keV. The reduced $\chi^2$ of the fit for each OBSID are generally  $\sim$ 0.8--1.5  with 62(63) dofs, showing mostly good fits.\footnote{  We have also checked the model COMPTT, but we generally derived a larger $\chi^2$, around 1.7 or beyond  with 62(63) dofs}.
The derived   DISKBB temperature ($T_{in}$) and the corresponding observed inner radius $R^{*}_{in}$
 of the emission area are shown in Table 1 and Fig. \ref{lc_spec}. Here,
$R^{*}_{in}$=$R_{in}\times(\cos\theta)^{1/2}$, where $R_{in}$=$D_{10}\times(N_{disk}/\cos\theta)^{1/2}$ is the inner radius of the disk,
$N_{disk}$ is the normalization on the disk blackbody component
via the DISKBB models assuming  $\theta=70^{\circ}$ and a distance
of 10 kpc, following McClintock et al. (2009).
The total flux in the energy band 1.5-100 keV is  also given in Table 1.
The  value of $L_{Edd}$ is $\sim$ $10^{39}$ erg s$^{-1}$ under an assumption of a black hole mass of 10 M$_{\odot}$ (McClintock et al. 2009).

At the initial part of the 2009 outburst, the flux increased in the first  6 OBSIDs (region I,  where the hardness decreases). The spectral fitting in this period shows that the temperature of the inner disk drops from 1.07 to 0.75 keV, whereas the radius of the inner disk increases from 5 km to 11 km, and the index of the power law spectrum steepens from 1.33 to 2.06, which are mostly cut off at energies around 60--80 keV (Table 1 and Fig. \ref{lc_spec}).
As the outburst decayed from the flux maximum (region II,  where the hardness remains at the lowest level, the peak luminosity is at $\sim$ 0.2 $L_{Edd}$), the  disk temperature decreases slowly from 0.85 keV to 0.66 keV, whereas the radius of the inner disk remains at $\sim$ 30 km and the  power-law index at $\sim$2.3. When the outburst  went into quiescence (region III,  where the hardness starts to increase again), the disk  temperature and the inner disk radius drop to 0.6 keV and  7 km, respectively.

The hardness at the left end of the HID is usually a diagnostic  of  the softest spectral state  that the  outburst reached.  Generally, the LHS has the largest hardness, with the SIMS and the HSS following it at decreasing hardness levels. Accordingly, the  hardness at the left end of HID usually corresponds to the  softest spectral state that the outburst reached.   McClintock et al. (2009) suggested a borderline  where the disk fraction is beyond 75\% to discriminate  the  HSS from the intermediate state.  To investigate this for the six outbursts of H~1743-322, we selected data at the left end of the HID for each outburst, and perform spectral  fits. The results are shown in Table 2. We find that only outbursts \#1 and \#2 have a disk fraction well beyond 0.75 (0.95$\pm0.01$ and 0.94$\pm0.02$, respectively),  \#3 and \#5 have it well below 0.75 (0.62$\pm0.01$ and 0.29$\pm0.02$, respectively), and the others (\#4 and \#6) have it around 0.75 (0.73$\pm0.02$ and 0.75$\pm0.02$, respectively). These results indicate that only the outbursts \#1 and \#2 have clearly  shown a HSS.

\begin{figure}[ptbptbptb]
\centering
 \includegraphics[angle=0, scale=0.9]{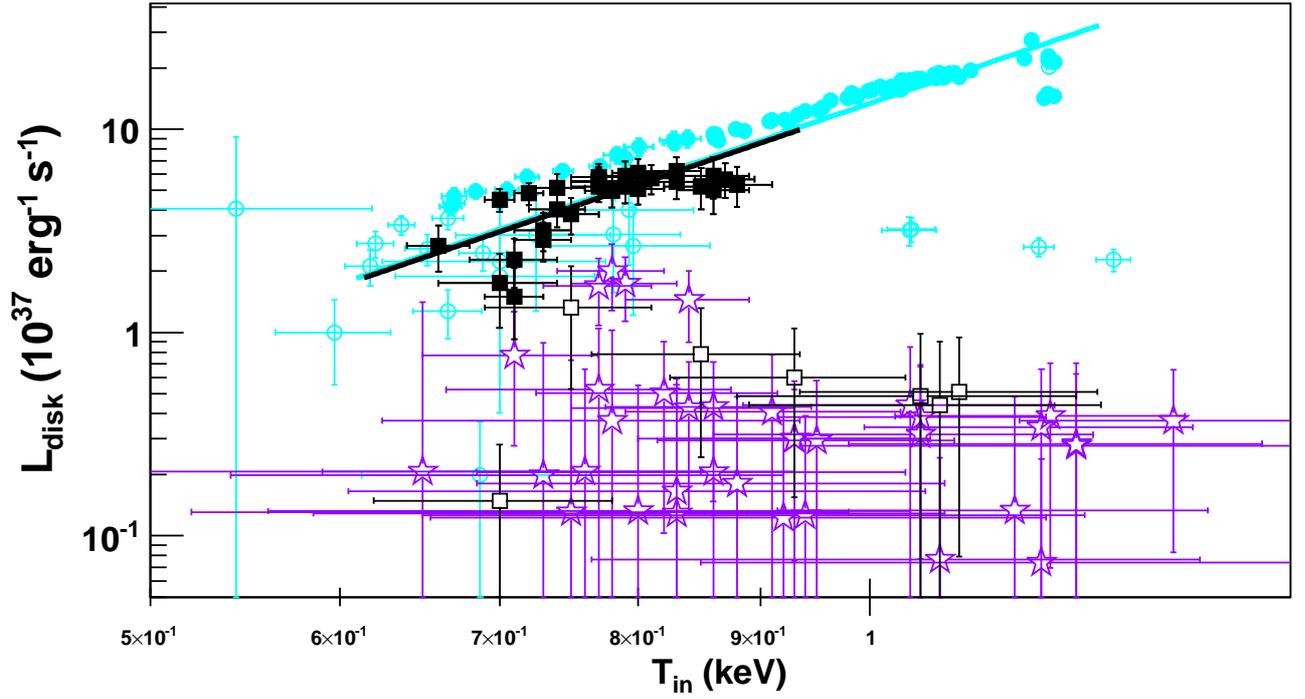}
      \caption{
        Bolometric disk luminosities $L_{disk}$   versus the temperature of the
inner disk $T_{in}$ for outbursts \#1 (light blue),  \#5 (violet), and
\#6 (black);
 filled markers for the thermal state OBSIDs, and open markers for the LHS OBSIDs.
  The fit  $L_{disk} \propto T_{in}^{4}$ is made in the thermal states
of the outbursts. The time bin size is one OBSID.
  The $T_{in}$--$L_{disk}$  relation for outburst \#1 was also
reported by McClintock et al. (2009).
       }
         \label{T_disk}
\end{figure}

It is believed that  in the  HSS of the BHC, $R_{in}$ remains quite constant
(e.g., see  Tanaka \& Lewin 1995 and McClintock et al. 2009).
The evidence for a constant inner disk radius in the HSS is that the bolometric luminosity of the disk, $L_{disk}$, is approximately proportional to the temperature of the inner disk $T_{in}$.
Assuming that the observed disk flux is  proportional to $\cos\theta$, we use $R_{in}$=$R_{in}^{*}/(\cos\theta)^{1/2}$ to calculate $L_{disk}$.
In Fig. \ref{T_disk}, we  show the fit of the relationship between bolometric luminosity of the disk and temperature of the inner disk $T_{in}$  in the HSS, with  $L_{disk} \propto T_{in}^{4}$.
The  reduced $\chi^{2}$ is  smaller in outburst 6 (1.3/27 dof) than in outburst 1 (40/99 dof),
with both showing deviations from the $T^{4}$ relationship. Such deviation was commonly observed in BH XRBs, for instance, in GRO J1655-40, XTE 1550-564,
and GRS 1915+105  (see the review by Tr\"umper \& Hasinger  2008). The possible explanation to this phenomenon might be the flux dependence of the spectral
hardening factor due to the Compton scattering of the soft photons in the corona region
\textbf{(Ohsuga et al. 2002, 2005; Kubota \& Makishima 2004; McClintock et al. 2009)}.
The HSS data of  the outbursts \#1 and \#6 at the HSS are the primary concerns of this plot.
For comparison, we  show the LHS data of outbursts \#1 and \#6,
and spectral data of the outburst \#5 as well.
 Note that for the outburst \#1, the data points were taken from Table 1 of McClintock et al. (2009). For outburst \#5, we extracted the spectra from every OBSIDs, since they were not available in Capitanio et al. (2009).
For the HSS of both outbursts, \#1 and \#6, Fig. \ref{T_disk} shows that the fit results are consistent with each other, which means that the inner disk  $R_{in}$ remained at a similar distance during  the HSS in these  two outbursts, despite of the large differences of the outburst evolution in flux and spectrum.

\begin{figure}[ptbptbptb]
\centering
 \includegraphics[angle=0, scale=0.8]{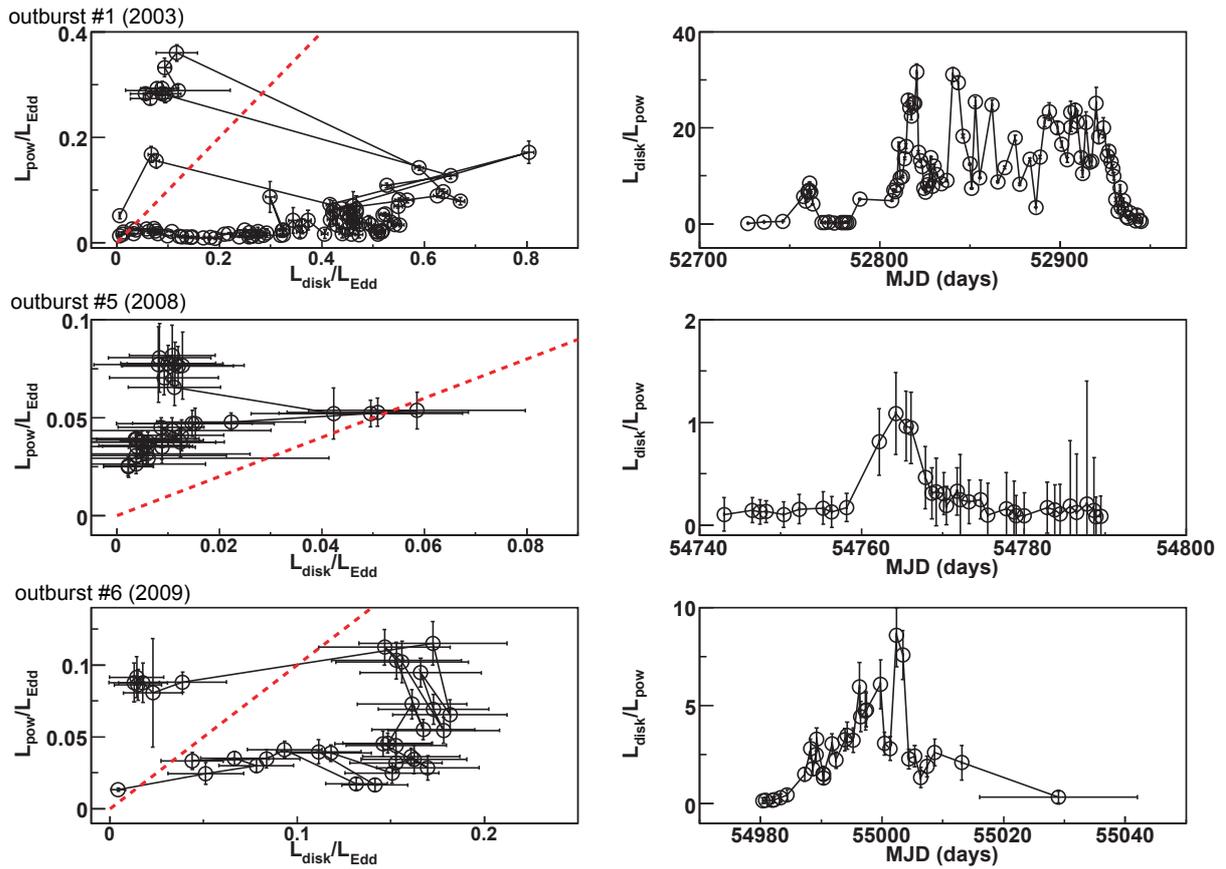}
      \caption{Left panels:  plot of the luminosity contribution of the power law versus the disk component in units of the Eddington luminosity, $L_{Edd}$.   The dashed  lines in the left panels show where the two components are equal.  Right panels:   time evolution of the ratio of the former contributions.
       }
         \label{disk_pow}
\end{figure}

The energy-division diagrams (see Remillard \& McClintock 2006), which were first used by Muno et al. (1999)
in describing the  behaviour of GRS~1915+105,  represent the  contributions of the   thermal and the non-thermal components.
Since the energy-division
diagrams are more effective than the HIDs in  disentangling intrinsic differences
between the states of BH binaries, we  show the Eddington ratios of the  disk bolometric  luminosity   versus the power-law luminosity
at 1-100 keV  (Fig. \ref{disk_pow}).
Here we consider only the outbursts \#1 (excluding  the VHS, defined by McClintock et al. 2009), \#5, and \#6, for which
whole  HID tracks were observed by  the PCA.
The dashed line in Fig. \ref{disk_pow} marks the spots where  the contribution from the  power-law  and disk fractions are equal. We find that any spectral state  beyond the LHS moves across this line, as is the case in the outbursts \#1 and \# 6. An additional support  of  this
argument comes from the energy-division diagram of the failed outburst \#5: the peak at the right end of the track  falls on the equal-division line.

\subsection{Hardness-RMS diagrams}

\begin{figure}[ptbptbptb]
\centering
 \includegraphics[angle=0, scale=0.4]{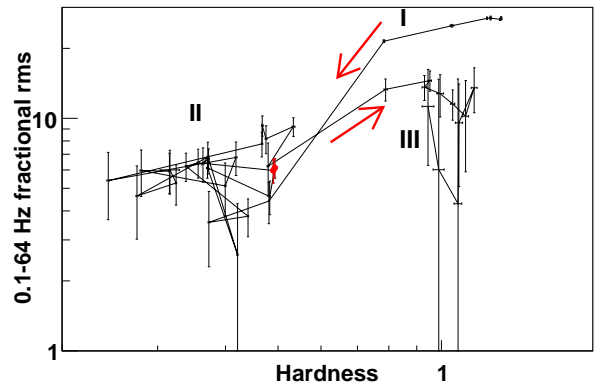}
      \caption{The hardness-rms diagrams (HRDs) of outburst \#6. The arrows show the evolution in time, and the three regions denote the LHS/HIMS (region I), the HSS (region II) and the LHS/HIMS (region III) during the decay of the outburst.  The two red points (almost on top of each other) are most likely corresponding a short SIMS. Each data point is derived from one OBSID.
}
         \label{lc_RMS}
\end{figure}

 For investigating the aperiodic variability of outburst \#6, we derived the lightcurves in time
bins of 1 ms (1/1024 s)  using the PCA Event mode data.
We adopted the method of Gleissner et al. (2004) to calculate the RMS by integrating the normalization PSDs in  the 0.1--64 Hz range.
The resulting track for the evolution of the RMS versus the hardness for outburst \#6 is plotted in Fig. \ref{lc_RMS}.
 As shown,  the HRD has a typical profile of a BHC.
During the LHS and HIMS (region I), the RMS is always above 20\%, and is positively correlated with the hardness.
After the  transition to the high state (region II), the RMS dropped to $\sim$5\% and
then moved to another LHS (region III, the LHS after the high state) with the RMS  values around 10\%.

 The PSD for all the OBSIDs in region I  show the band-limited noise and type-C QPOs at $\sim$1~Hz, typical for LHS and HIMS.
No QPO is detected in the other OBSIDs in the 0.01--500~Hz band, except for two OBSIDs (94413-01-03-02 \& 94413-01-03-03), which are marked with red points in Fig. \ref{lc_RMS}), with type-B QPOs found at $\sim$4~Hz in region II. As was shown already in the outbursts of other  sources (e.g., Belloni 2009), this resembles the SIMS, which is usually located in  the HRD within a small region beneath the transition   from  the HIMS to the HSS along the overall evolution. Note that the following OBSID 94413-01-03-04 (MJD 54990.33)  was carried out half an hour latter, with no QPO detected. If the SIMS is correlated with a jet as suggested in Belloni et al. (2005) and Fender et al. (2006),  this may indicate short-lived   jitters on timescales of around 1 hour.
All the other  OBSIDs with the lower hardness  in region II of the HRD have  RMS values of only a few per cent, no QPO, and a relatively larger disk fraction (at most around 75 $\%$). These results,  together with the constant inner disk radius, indicate that  the outburst \#6 most likely reached the HSS, although  not at  the leftmost region of the HID as in the case of the 2003 outburst, and with a disk fraction being just  near the borderline of the canonical HSS.

\section{Discussion and summary}

We have analyzed the 2009 outburst from H~1743-322 and made comparisons with the other five outbursts as observed by RXTE since 2003.
We have found that the 2009 outburst of H~1743-322  is different to the other five outbursts observed to date. The HID shows the third complete counter-clockwise q-track  for this source during the 2009 outburst. Interestingly, this track is located between  the huge one in 2003 --which has a
complete state transition to HSS-- and the prior outburst from the source in 2008, which is a failed outburst. In the HID, the 2009 outburst leaves the LHS but does not reach the leftmost edge attained by the 2003 outburst. The softest point in HID has  only intermediate hardness values of 0.2--0.3, with X-ray spectra described by a power law with photon index  $<$ 2.4, and a peak luminosity at $\sim$ 0.2 $L_{Edd}$.  This behaviour is similar to that of  a ``failed
state transition" seen in the  persistent bright black hole binary Cyg~X-1 (Belloni  2009).  Cyg~X-1 is not a transient black hole system: it is found most of the time in the LHS with occasional transitions to the HSS, and a series of ``failed state transitions" (see e.g., Fig. 8 in Belloni 2009).
The ``failed state transitions" only reach the
SIMS with  the X-ray spectra described by a power law with a photon index of $\sim$ 2.15 and  an inner disk temperature of $\sim$ 0.36 keV. For H~1743-322, the leftmost edge in HID of the 2009 outburst has a disk fraction of 75$\%$, which is close to the lower limit observed for the XRB sample which complete the state transitions to the HSS (McClintock et al. 2009).
 The timing analysis shows, however, typical features of reaching a HSS. Thus   this outburst may be regarded as an interesting event: one in which   the  ``failed state transitions"  can just be surpassed to reach the HSS. Therefore, H1743-322 could have a failed outburst at low flux and hardness level (as indicated in the 2008 outburst), and high soft  states with a rather wide range in the flux and the hardness values (as indicated  in the  2003 and 2009 outbursts).

Advection-dominated accretion flow (ADAF) models (e.g. Esin et al.
1997) predict that the inner radius of the accretion disk should be radially truncated at hundreds of gravitational radii in the LHS
and the disk should extend to the ISCO  in the HSS.
Indeed,
GX~339-4 (Motta et al. 2009) and some other BHC follow this phenomenology.
However, it is interesting to note that in the latest outburst of H~1743-322,
the accretion disk is closer to the innermost stable circular orbit in LHS rather than in softer state.
Similar conclusions based on the apparent
change in the normalization of the DISKBB model have been
presented by Belloni et al. (1997) for the black hole binary GRS~
J1915+105 in the LHS using RXTE/PCA data,
and Miller et al. (2006) for Swift~J1753.5-0127 using XMM-Newton data.
Reis et al. \textbf{(2010)} investigated 8 stellar mass black
holes in LHS in the 0.5-100 keV broad band energy range using observations from XMM-Newton, Suzaku, Chandra, Swift and RXTE: their results also suggested that the disk in the LHS can remain at
the ISCO, rather than being truncated at hundreds of gravitational radii.
A similarity of  the outbursts from H~1743-322 with those seen in a typical black hole transient like GX 339-4 is that the most strong outburst was followed by a sequence of weaker ones. A cool truncated accretion disk can be left over at around  ISCO from the strongest outburst,
and shows up in the low hard state of the following weaker ones (Meyer-Hofmeister et al. 2009).

We find that for the 2009 outburst, the inner radius of the accretion disk has an anti-correlation with disk temperature.
Such a trend was present as well  in the 2003 outburst (McClintock et al. 2009). One possible  understanding of this feature may relate to
 a reprocess of the hot corona on the soft X-rays from the accretion disk, which can lead to an underestimation of the inner radius of the accretion disk. By considering that the Comptonized component illuminate the disk, the irradiated disk can have a higher temperature and a smaller inner disk radius than a non-irradiated disk
(e.g. Yao et al. 2001, Gierli\`nski et al. 2008).
Gierli\`nski et al. (2008) suggest that the disk spectra are largely   unaffected in the HSS, in which the power-law flux is less than that of the disk, and it does not hold for   LHS with $L_{pow}$/$L_{disk}$ $>$ 1.
 The underestimation factor of the inner radius depends on the strength of $L_{pow}/L_{disk}$ (see Fig. 6 in Gierli\`nski et al. 2008). For example,
in  the case $L_{pow}$/$L_{disk}$ =5,  the DISKBB  will model the inner disk radius with  a factor $\sim$ 4 lower.
In the LHS of the 2009 outburst from  H~1743-322 we have  $L_{pow}$/$L_{disk}$$>$10. Therefore the inner disk  could be underestimated by at least  a factor $>$ 5. By taking this into account, the inner radius of the accretion at LHS should be comparable or larger than that in   other spectral states.

The transition from the low hard state to other states is usually accompanied with a balance between the thermal and the non-thermal emissions.
We find for the first time that such a balance might account   for  the 2008 failed outburst of H 1743-322.
As shown in Fig. 6,  the tracks of the 2003 and 2009 outbursts go well across the so-called equal-division line, which represents equal  thermal and non-thermal  contributions to the overall observed emission, to leave LHS for further states. Instead, the track of the 2008 outburst has a turn-over  on top of the  equal-division line. This may be understood if the shower of the soft X-ray emission from the accretion disk cool  the corona. The   equal-division line in Fig. 6 may be  indicative of
emission balance between the corona and the inner accretion disk, beyond which
  the spectral state will go further than the LHS.

In summary,  we have investigated the spectral and temporal evolution of the 2009  outburst, presenting the third complete q-track  from H~1743-322.
 We found that the 2009 outburst can reach the HSS by just overcoming  a  ``failed state transition". This finding, together with the other two complete q-tracks from the source, showing a HSS at rather low hardness in 2003,  and a failed outburst in 2008, may constitute an interesting testbed for further investigating  the evolution of the spectral states in BH XRBs.

\acknowledgements
This work was subsidized by the National Natural Science Foundation of China, the CAS key Project KJCX2-YW-T03, and 973 program 2009CB824800. J.-M. W. thanks the Natural Science Foundation of China for support via NSFC-10325313, 10521001 and 10733010. DFT acknowledges support from the grants AYA2009-07391 and SGR2009-811. We thank an anonymous referee for her/his careful comments on the first version of the manuscript.

\onecolumn

\begin{table}[ptbptbptb]
\begin{center}
\label{tab_outburst6}
\caption{The spectral fit results of the outburst in 2009.
 }
\begin{tabular}{cccccccccccccccccc}
\hline \hline
 No & OBSID & Time & $T_{in}$   &$R^{*}_{in}$  &$\Gamma_{1}$  &N    &$E_{cut}$       &Flux&$\chi_{red}^{2}(d.o.f)$ \\\hline
    &        & (MJD) &keV         &km       &           & &keV             &$10^{-9}$  &\\\hline
1 &94413-01-02-00  &54980.40  &$1.07_{-0.14}^{+0.22}$  &$5_{-1}^{+3}$  &$1.33_{-0.09}^{+0.07}$  &$0.26_{-0.04}^{+0.04}$  &$62.00_{-12.00}^{+16.00}$   &7.02$\pm$0.15   &0.90(62) \\\hline
2 &94413-01-02-02  &54980.85  &$1.09_{-0.12}^{+0.19}$  &$5_{-1}^{+2}$  &$1.30_{-0.07}^{+0.06}$  &$0.24_{-0.03}^{+0.03}$  &$59.00_{-8.00}^{+11.00}$   &7.33$\pm$0.12   &0.96(62) \\\hline
3 &94413-01-02-01  &54981.96  &$1.05_{-0.14}^{+0.20}$  &$6_{-2}^{+3}$  &$1.40_{-0.09}^{+0.08}$  &$0.32_{-0.05}^{+0.05}$  &$67.00_{-14.00}^{+22.00}$   &7.32$\pm$0.18   &0.88(62) \\\hline
4 &94413-01-02-05  &54982.28  &$0.93_{-0.10}^{+0.11}$  &$8_{-2}^{+3}$  &$1.42_{-0.08}^{+0.08}$  &$0.32_{-0.05}^{+0.05}$  &$69.00_{-13.00}^{+18.00}$   &7.30$\pm$0.13   &0.94(62) \\\hline
5 &94413-01-02-04  &54983.33  &$0.85_{-0.08}^{+0.09}$  &$11_{-2}^{+4}$  &$1.63_{-0.07}^{+0.07}$  &$0.49_{-0.24}^{+0.22}$  &$81.00_{-18.00}^{+31.00}$   &6.88$\pm$0.13   &1.20(62) \\\hline
6 &94413-01-02-03  &54984.38  &$0.75_{-0.06}^{+0.06}$  &$18_{-4}^{+6}$  &$2.06_{-0.03}^{+0.03}$  &$1.14_{-0.09}^{+0.09}$  &   &7.47$\pm$0.12   &1.21(63) \\\hline
7 &94413-01-03-00  &54987.26  &$0.86_{-0.03}^{+0.03}$  &$29_{-2}^{+3}$  &$2.24_{-0.05}^{+0.05}$  &$2.15_{-0.28}^{+0.29}$  &   &13.40$\pm$0.19   &1.29(63) \\\hline
8 &94413-01-03-01  &54988.24  &$0.83_{-0.02}^{+0.02}$  &$32_{-2}^{+2}$  &$2.24_{-0.06}^{+0.06}$  &$1.22_{-0.19}^{+0.20}$  &   &9.97$\pm$0.16  &0.99(63) \\\hline
9 &94413-01-03-07  &54988.63  &$0.87_{-0.02}^{+0.03}$  &$28_{-2}^{+2}$  &$2.27_{-0.04}^{+0.03}$  &$1.87_{-0.20}^{+0.20}$  &   &11.60$\pm$0.18   &0.96(63) \\\hline
10 &94413-01-03-05  &54989.16  &$0.79_{-0.02}^{+0.02}$  &$34_{-2}^{+3}$  &$2.31_{-0.08}^{+0.07}$  &$1.47_{-0.21}^{+0.23}$  &   &8.65$\pm$0.17  &0.85(63) \\\hline
11 &94413-01-03-06  &54989.22  &$0.80_{-0.02}^{+0.02}$  &$34_{-2}^{+2}$  &$2.22_{-0.07}^{+0.06}$  &$0.98_{-0.16}^{+0.17}$  &   &8.84$\pm$0.16   &0.76(63) \\\hline
12 &94413-01-03-02  &54990.27  &$0.88_{-0.03}^{+0.03}$  &$26_{-2}^{+3}$  &$2.21_{-0.06}^{+0.05}$  &$1.80_{-0.25}^{+0.26}$  &   &12.00$\pm$0.18  &1.40(63) \\\hline
13 &94413-01-03-03  &54990.33  &$0.86_{-0.03}^{+0.03}$  &$27_{-2}^{+3}$  &$2.28_{-0.04}^{+0.04}$  &$2.26_{-0.25}^{+0.25}$  &   &12.20$\pm$0.19   &0.97(63) \\\hline
14 &94413-01-03-04  &54990.39  &$0.85_{-0.03}^{+0.03}$  &$28_{-2}^{+3}$  &$2.23_{-0.05}^{+0.05}$  &$1.89_{-0.23}^{+0.24}$  &   &11.97$\pm$0.19   &1.17(63) \\\hline
15 &94413-01-03-08  &54991.77  &$0.81_{-0.02}^{+0.02}$  &$32_{-2}^{+2}$  &$2.21_{-0.05}^{+0.05}$  &$0.97_{-0.12}^{+0.13}$  &   &8.60$\pm$0.15   &1.12(63) \\\hline
16 &94413-01-03-09  &54992.43  &$0.83_{-0.02}^{+0.02}$  &$30_{-2}^{+2}$  &$2.26_{-0.06}^{+0.06}$  &$1.41_{-0.19}^{+0.21}$  &   &9.74$\pm$0.16   &1.10(63) \\\hline
17 &94413-01-03-10  &54993.86  &$0.80_{-0.02}^{+0.02}$  &$31_{-2}^{+2}$  &$2.14_{-0.06}^{+0.05}$  &$0.70_{-0.10}^{+0.11}$  &   &7.52$\pm$0.13   &1.07(63) \\\hline
18 &94413-01-04-00  &54994.25  &$0.79_{-0.02}^{+0.02}$  &$32_{-2}^{+2}$  &$2.15_{-0.09}^{+0.08}$  &$0.69_{-0.15}^{+0.17}$  &   &7.49$\pm$0.14   &1.44(63) \\\hline
19 &94413-01-04-01  &54995.24  &$0.78_{-0.02}^{+0.02}$  &$32_{-2}^{+3}$  &$2.18_{-0.08}^{+0.08}$  &$0.75_{-0.14}^{+0.16}$  &   &7.26$\pm$0.14  &0.90(63) \\\hline
20 &94413-01-04-03  &54996.28  &$0.77_{-0.02}^{+0.02}$  &$36_{-2}^{+2}$  &$2.12_{-0.11}^{+0.01}$  &$0.42_{-0.11}^{+0.14}$  &   &6.71$\pm$0.14   &1.18(63) \\\hline
21 &94413-01-04-02  &54996.48  &$0.77_{-0.02}^{+0.02}$  &$35_{-2}^{+2}$  &$2.16_{-0.07}^{+0.06}$  &$0.58_{-0.09}^{+0.10}$  &   &6.86$\pm$0.13  &1.39(63) \\\hline
22 &94413-01-04-08  &54997.27  &$0.77_{-0.02}^{+0.02}$  &$35_{-2}^{+3}$  &$2.17_{-0.11}^{+0.10}$  &$0.56_{-0.15}^{+0.17}$  &   &6.75$\pm$0.14   &1.74(63) \\\hline
23 &94413-01-04-05  &54997.40  &$0.77_{-0.02}^{+0.02}$  &$34_{-2}^{+2}$  &$2.15_{-0.08}^{+0.08}$  &$0.50_{-0.10}^{+0.10}$  &   &6.35$\pm$0.12   &1.20(63) \\\hline
24 &94413-01-04-06  &54999.75  &$0.74_{-0.02}^{+0.02}$  &$36_{-2}^{+2}$  &$2.14_{-0.11}^{+0.10}$  &$0.38_{-0.10}^{+0.12}$  &   &5.74$\pm$0.12  &1.29(63) \\\hline
25 &94413-01-04-04  &55000.40  &$0.74_{-0.02}^{+0.02}$  &$32_{-2}^{+3}$  &$2.17_{-0.06}^{+0.06}$  &$0.63_{-0.09}^{+0.10}$  &   &5.91$\pm$0.12  &0.72(63) \\\hline
26 &94413-01-05-00  &55001.32  &$0.75_{-0.02}^{+0.02}$  &$31_{-2}^{+3}$  &$2.18_{-0.09}^{+0.08}$  &$0.66_{-0.13}^{+0.15}$  &   &5.72$\pm$0.12   &1.05(63) \\\hline
27 &94413-01-05-01  &55002.36  &$0.72_{-0.01}^{+0.01}$  &$37_{-2}^{+2}$  &$1.96_{-0.12}^{+0.11}$  &$0.17_{-0.04}^{+0.05}$  &   &4.68$\pm$0.10   &1.06(63) \\\hline
28 &94413-01-05-02  &55003.41  &$0.70_{-0.01}^{+0.01}$  &$38_{-2}^{+2}$  &$2.05_{-0.09}^{+0.06}$  &$0.22_{-0.05}^{+0.05}$  &   &4.53$\pm$0.10   &1.06(63) \\\hline
29 &94413-01-05-03  &55004.39  &$0.73_{-0.02}^{+0.02}$  &$29_{-2}^{+3}$  &$2.22_{-0.03}^{+0.05}$  &$0.74_{-0.10}^{+0.11}$  &   &5.28$\pm$0.11   &1.10(63) \\\hline
30 &94413-01-05-04  &55005.37  &$0.73_{-0.02}^{+0.02}$  &$28_{-2}^{+3}$  &$2.13_{-0.07}^{+0.07}$  &$0.52_{-0.09}^{+0.10}$  &   &4.56$\pm$0.09  &0.86(63) \\\hline
31 &94413-01-05-05  &55006.36  &$0.71_{-0.02}^{+0.02}$  &$21_{-3}^{+4}$  &$2.14_{-0.07}^{+0.07}$  &$0.51_{-0.09}^{+0.09}$  &   &4.38$\pm$0.09   &1.01(63) \\\hline
32 &94413-01-05-06  &55007.40  &$0.71_{-0.03}^{+0.03}$  &$26_{-2}^{+3}$  &$2.15_{-0.06}^{+0.05}$  &$0.55_{-0.07}^{+0.08}$  &   &4.12$\pm$0.08   &1.27(63) \\\hline
33 &94413-01-06-00  &55008.65  &$0.66_{-0.02}^{+0.02}$  &$33_{-3}^{+4}$  &$2.22_{-0.07}^{+0.06}$  &$0.54_{-0.08}^{+0.09}$  &   &3.82$\pm$0.10  &1.00(63) \\\hline
34 &94413-01-06-01  &55013.09  &$0.70_{-0.04}^{+0.04}$  &$24_{-3}^{+5}$  &$2.02_{-0.13}^{+0.12}$  &$0.29_{-0.08}^{+0.10}$  &   &3.00$\pm$0.09  &1.05(63) \\\hline
35 &after 94413-01-07-00  &55029.00  &$0.70_{-0.08}^{+0.08}$  &$7_{-2}^{+4}$  &$1.87_{-0.04}^{+0.04}$  &$0.11_{-0.01}^{+0.01}$  &   &1.04$\pm$0.02   &1.18(63) \\\hline

\end{tabular}
\end{center}
\begin{list}{}{}
\item[Note:]{
RXTE/PCA and RXTE/HEXTE data are used for the spectral fit. The OBSID, time,  the inner temperature of disk, the inner radius of the disk, power law index, break for the energy, and unabsorbed flux in the energy band 1.5--100 keV (in units of $10^{-9}$ erg~cm$^{-2}$ s$^{-1}$) are given. PCA spectra are derived for each  OBSID ($\sim$ 3000 seconds) and from the combining OBSIDs for the last 11 data segments.  Here
$R^{*}_{in}$=$R_{in}\times(\cos\theta)^{1/2}$ and $R_{in}$=$D_{10}\times(N_{disk}/\cos\theta)^{1/2} $ is the inner radius of the disk, $\theta$ is the inclination angle, $N_{disk}$ is the normalization on the disk blackbody component, $D_{10}$ is the distance of the source in units of 10 kpc.
}
\end{list}
\end{table}

\begin{table}[ptbptbptb]
\begin{center}
\label{tab_outburst}
\caption{The spectral fit results of the softest OBSID during the six outbursts.
 }
\begin{tabular}{cccccccccccccccccc}
\hline \hline
 No & OBSID & Time & $T_{in}$   &$R^{*}_{in}$  &$\Gamma_{1}$  &N           &Flux&$\chi_{red}^{2}(d.o.f.)$ &$F_{disk}/Flux$\\\hline
    &        & (MJD) &keV         &km                 & &             &$10^{-9}$  & &\\\hline
1 &80146-01-15-10  &52900.87  &$0.91_{-0.01}^{+0.02}$  &$32_{-1}^{+1}$  &$2.35_{-0.41}^{+0.28}$  &$0.14_{-0.09}^{+0.15}$      &8.84$\pm$0.13   &0.92(36)  & 0.95$\pm$0.01\\\hline
2 &90058-10-01-00  &53197.32  &$0.85_{-0.01}^{+0.01}$  &$32_{-1}^{+1}$  &$2.58_{-0.24}^{+0.24}$  &$0.21_{-0.10}^{+0.17}$     &6.38$\pm$0.13
&0.48(36)  &0.94$\pm$0.02\\\hline
3 &91050-06-05-00  &53604.50  &$0.83_{-0.02}^{+0.02}$  &$29_{-1}^{+1}$  &$2.28_{-0.04}^{+0.07}$  &$0.83_{-0.12}^{+0.14}$      &7.40$\pm$0.16   &0.79(63) &0.62$\pm$0.01\\\hline
4 &93427-01-01-00  &54481.98  &$0.74_{-0.02}^{+0.02}$  &$34_{-2}^{+2}$  &$1.93_{-0.05}^{+0.06}$  &$0.16_{-0.05}^{+0.06}$      &4.47$\pm$0.11   &0.97(63) &0.73$\pm$0.02\\\hline
5 &93427-01-12-00  &54764.28  &$0.78_{-0.04}^{+0.04}$  &$20_{-2}^{+3}$  &$2.14_{-0.07}^{+0.06}$  &$0.86_{-0.14}^{+0.16}$      &5.64$\pm$0.11   &0.91(63) & 0.29$\pm$0.02\\\hline
6 &94413-01-05-01  &55002.36  &$0.72_{-0.01}^{+0.01}$  &$37_{-2}^{+2}$  &$1.96_{-0.12}^{+0.11}$  &$0.17_{-0.04}^{+0.05}$      &4.68$\pm$0.10
&1.06(63) & 0.75$\pm$0.02\\\hline
\end{tabular}
\end{center}
\begin{list}{}{}
\item[Note:]{
RXTE/PCA and RXTE/HEXTE data are used except  for the first 2 outbursts. The OBSID, time, the inner temperature of disk, the inner radius of the disk, power law index, break for the energy, unabsorbed flux in the energy band 1.5--100 keV in units of $10^{-9}$ erg cm$^{-2}$ s$^{-1}$, and the fraction of DISKBB component are given.
We notice that the spectral results of OBSID 80146-01-15-10 and of OBSID 93427-01-12-00 were  also reported by McClintock et al. (2009) and Capitanio et al (2009), respectively.
}
\end{list}
\end{table}

\end{document}